\documentclass[12pt]{article}
\usepackage{graphicx}
\usepackage[cp1251]{inputenc}
\usepackage{epsfig}
\usepackage[english]{babel}

\date{}

\begin{document}
\setcounter{page}{1}
\pagestyle{plain}

\title{\bf{Electronic spectrum and optical properties of Y-shaped Kekul\'{e}-patterned graphene: Band nesting resonance as an optical signature}}

\author{Yawar Mohammadi \thanks{E-mail address:
y.mohammadi@cfu.ac.ir}}
\maketitle{\centerline{Department of Physics, Farhangian University, Tehran, Iran}

\begin{abstract}

Employing tight-binding model we investigate the effects of a uniform Y-shaped Kekul\'{e} lattice distortion on the electronic spectrum and optical conductivity of graphene. We derive a low-energy effective Hamiltonian which is found to be in excellent agreement with one calculated from a diagonalization of the full tight-binding Hamiltonian. Then using the low-energy Hamiltonian and Kubo formula we obtain an analytical expression for the real part of the optical conductivity used to explore the effects of chemical potential, temperature and on-site and hopping energy deviations in details. In particular we find that Y-shaped Keku\'{e}-patterned graphene at finite chemical potential displays a large optical response called \textit{band nesting resonance}. This effect is shown to be robust against increasing temperature, facilitating its detection as an optical signature for the Y-shaped Kekul\'{e} distortion even at room temperature.

\end{abstract}


\vspace{0.5cm}

\section{Introduction}
\label{sec01}

Recently Guti\'{e}rrez \textit{et. al} \cite{Gutierrez1} have realized experimentally a Y-shaped periodic alternation of weak and strong C-C bonds in a superlattice of graphene grown epitaxially onto copper substrate, called Y-shaped Kekul\'{e} lattice distortion with reference to the Kekul\'{e} dimerization in a benzene ring. Using density functional theory calculations they showed \cite{Gutierrez1} that removing copper atoms from the topmost surface allows the system to relax created an alternating network of C-C bonds similar to the Y-shaped Kekul\'{e} lattice distortion. Gamayun \textit{et.al} introduced a tight-binding Hamiltonian to explore the effects of a uniform Y-shaped Kekul\'{e} lattice on the properties of graphene, in which this effect has been taken into account by adding a nearest-neighbor hopping energy deviation to the minimal tight-binding Hamiltonian of graphene. According to this Hamiltonian the uniform Y-shaped Kekul\'{e} distortion locks the valley degree of freedom of the charge carriers to their direction of motion, resulting in the breaking of the valley degeneracy of graphene and emerging two species of massless Dirac fermions\cite{Gamayun1}. Some researchers, using this Hamiltonian, have investigated different properties of Y-shaped Kekul\'{e}-patterned graphene. Wu \textit{et. al}\cite{Wu1} proposed a type of valley field-effect transistors for Y-shaped Kekul\'{e}-patterned graphene and explored tuning valley pseudomagnetoresistance via an electric field.  Andrade \textit{et. al}\cite{Andrade1} studied the effects of uniaxial strain on the band structure of all types of Kekul\'{e}-distorted graphene. In other works the effect of the Y-shaped Kekul\'{e} distortion on the electronic transport properties\cite{Andrade2,Wang1}, dynamical polarization\cite{Herrera1}, magneto-optical conductivity\cite{Mohammadi1} and quantum Hall effect\cite{Mohammadi2} in graphene has been investigated.

In this paper we consider the electronic spectrum and the optical conductivity of such a system with particular emphasis on the effects of the on-site energy deviation, which in the previous researches has been neglected\cite{Wu1,Andrade1,Andrade2,Wang1,Herrera1,Mohammadi1,Mohammadi2,Mojarro1,Herrera2} while as shown here it leads to fascinating optical properties. The structure of our paper is organized as follows. In section \ref{sec02}, we introduce our tight-binding model in which the effects of the Y-shaped Kekul\'{e} distortion is taken into account by including both on-site and hopping energy deviations in the minimal tight-binding Hamiltonian of graphene. Then we derive a low-energy effective Hamiltonian in subsection \ref{sec02-1} which is found to be in excellent agreement with one calculated from a diagonalization of the full tight-binding Hamiltonian. The main effect of the on-site energy deviation on the band structure is that a set of bands gains an effective mass and a shift in energy, thus lifting the degeneracy of the conduction bands at the Dirac point. In the next subsection we obtain an analytical expression for the real part of the optical conductivity using Kubo formula, which is used to explore the effects of temperature, chemical potential and on-site and nearest-neighbor hopping energy deviations on the optical conductivity. Then we present our results for the optical conductivity as a function of the photon energy in next section. In particular we find that in the limit of zero chemical potential the optical conductivity displays a dip-peak structure located at the photon energy corresponding to 2 times the effective on-site energy deviation. Furthermore, it is shown that at finite chemical potential Y-shaped Keku\'{e}-patterned graphene exhibits a large optical response caused by nesting of the conduction or valance bands. This effect is shown to be robust with respect to increasing temperature, facilitating its detection even at room temperature. In this section we also discuss the effects of next-nearest-neighbor hopping energy deviation. Finally we end the paper by presenting summary and conclusions in section \ref{sec04}.

\section{Model and Formulation}
\label{sec02}

A schematic representation of Y-shaped Kekul\'{e}-patterned graphene superlattice has been displayed in Fig. \ref{Fig01}(a), in which the three thick lines in each unit cell of graphene superlattice denote the bonds that connect the carbon atom located on the copper-atom vacancies in substrate to its nearest neighbors. These bonds acquire a shorter nearest-neighbor bond~\cite{Gutierrez1}, leading to elongation of the other nearest-neighbor bonds, which is shown by thin lines. Also the on-site energy of carbon atoms located on the cooper-atom vacancies is expected to be changed with respect to other carbon atoms. Therefore, by taking into account deviations in the hopping and on-site energies~\cite{Harrison1}, the tight-binding Hamiltonian for Y-shaped Kekul\'{e}-patterned graphene in the nearest-neighbor approximation is given by

\begin{eqnarray}\label{eq01}
H&=&u\sum_{i=1}^{N/3}a_{\vec{\mathbf{R}}_{i}}^{\dag}a_{\vec{\mathbf{R}}_{i}}-(t_{0}+\delta t) \sum_{i=1}^{N/3}\sum_{n=1}^{3}(b_{\vec{\mathbf{R}}_{i}+\vec{\mathbf{\delta}}_{n}}^{\dag}a_{\vec{\mathbf{R}}_{i}}+h.c.) \nonumber \\
&-&(t_{0}-\delta t)\sum_{i=1}^{N/3}\sum_{n=1}^{3}(b_{\vec{\mathbf{R}}_{i}
+\vec{\mathbf{\delta}}_{3}-\vec{\mathbf{\delta}}_{1}+\vec{\mathbf{\delta}}_{n}}^{\dag}
a_{\vec{\mathbf{R}}_{i}+\vec{\mathbf{\delta}}_{3}-\vec{\mathbf{\delta}}_{1}}
+b_{\vec{\mathbf{R}}_{i}+\vec{\mathbf{\delta}}_{3}-\vec{\mathbf{\delta}}_{2}+\vec{\mathbf{\delta}}_{n}}^{\dag}a_{\vec{\mathbf{R}}_{i}+
\vec{\mathbf{\delta}}_{3}-\vec{\mathbf{\delta}}_{2}}+h.c.)
\end{eqnarray}
where $t_{0}$ and $N$ are the nearest-neighbor hopping energy and the number of the unit cells in pristine graphene respectively (The number of the nuit cells of the Y-shaped Kekul\'{e}-patterned graphene is $N/3$.). The operator $a_{\vec{\mathbf{R}}_{i}}^{\dag}$ ($b_{\vec{\mathbf{R}}_{i}}^{\dag}$) creates an electron in the carbon atoms on sublattice $A$ ($B$) located at $i^{th}$ unit cell. $\vec{\mathbf{R}}_{i}=n_{i}\vec{\mathbf{A}}_{1}+m_{i}\vec{\mathbf{A}}_{2}$ are the translation vectors of Y-shaped Kekul\'{e}-patterned graphene with $\vec{\mathbf{A}}_{1}=(3\sqrt{3}a/2,3a/2)$ and $\vec{\mathbf{A}}_{2}=(-3\sqrt{3}a/2,3a/2)$ its primitive translation vectors, and $n_{i}$ and $m_{i}$ are integer numbers. The vectors $\vec{\mathbf{\delta}}_{1}=(\sqrt{3}a/2,a/2)$, $\vec{\mathbf{\delta}}_{2}=(\sqrt{3}a/2,a/2)$ and $\vec{\mathbf{\delta}}_{3}=(0,-a)$ are drown from each A carbon atom to its nearest neighbors and $a$ is the shortest carbon-carbon distance. The on-site energy deviation, $u$, is assumed to be nonzero only for the carbon atoms located on the cooper-atom vacancies in the substrate. The hopping energy deviation, which is due to the change in the bond length, is assumed to be $+\delta t$ and $-\delta t$ for carbon-carbon bonds shown by thick and thin lines, respectively, in Fig. \ref{Fig01}(a). This Hamiltonian can be diagonalized by selecting the appropriate unit cell, red hexagonal in \ref{Fig01}(a), in space k. The corresponding band structure for different values of the hopping and on-site energy deviations, (a) $\delta t=0.1t_{0}$ and $u=0$, (b) $\delta t=0$ and $u=0.1t_{0}$, (c) $\delta t=0.1t_{0}$ and $u=0.1t_{0}$ has been shown in Figure \ref{Fig01} by solid black lines.

\begin{figure}
\begin{center}
\includegraphics[width=17.5cm,angle=0]{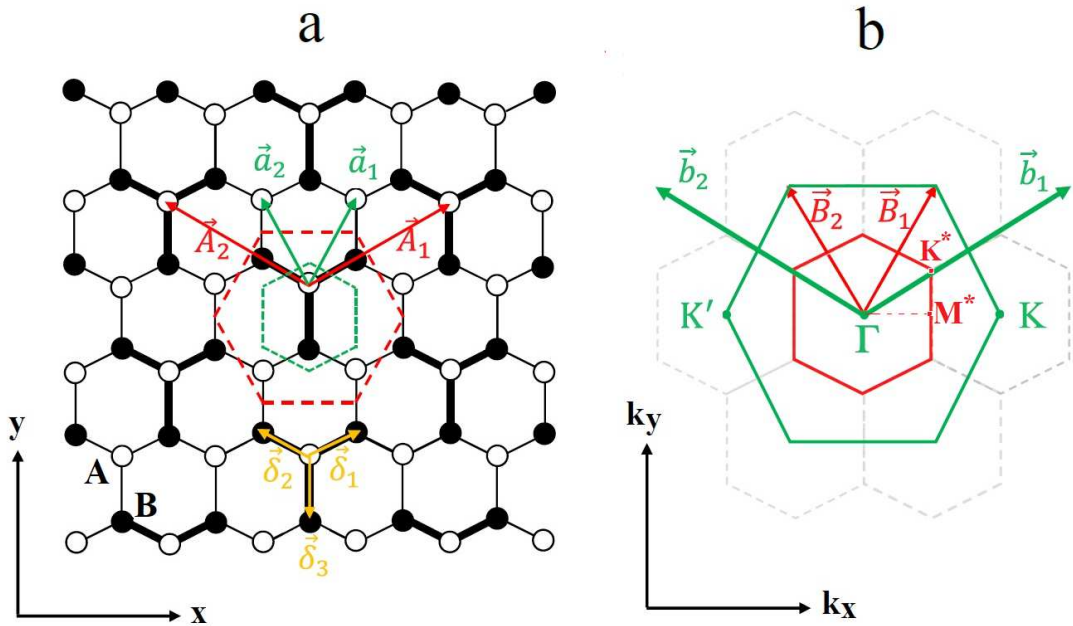}
\caption{(a) A schematic representation of a single layer graphene with a uniform Y-shaped Kekul\'{e} lattice distortion in which primitive cells of pristine  and Y-shaped Kekul\'{e}-patterned has been shown by green and red hexagonal respectively. $\vec{a}_{i}$ ($\vec{A}_{i}$) indicates primitive translation vectors of pristine (Y-shaped Kekul\'{e}-patterned) graphene. (b) Reciprocal lattices for pristine(in green) and Y-shaped Kekul\'{e}-patterned (in black) graphene.}
\label{Fig01}
\end{center}
\end{figure}

\subsection{Low-energy effective Hamiltonian}
\label{sec02-1}

To build a low-energy effective Hamiltonian, we must first express the Hamiltonian of Eq. 1 in the momentum space. As it is clear from Fig. \ref{Fig01}(b), the first Brillouin zone of pristine graphene can be represented through three copies of that of graphene superlattice which are set to be centered at $\mathbf{K}_{j}$ (j=1,2 and 3), $\mathbf{K}_{1}=\mathbf{K}=(+4\pi/3\sqrt{3},0)$, $\mathbf{K}_{2}=\mathbf{\Gamma}=(0,0)$ and $\mathbf{K}_{2}=\mathbf{K}^{'}=(-4\pi/3\sqrt{3},0)$.  Therefore, the annihilation (creation) operators can be expanded in the momentum space as \cite{Ren1} $a_{i}=\frac{1}{\sqrt{N}}\sum_{\mathbf{k}}\sum_{j}e^{i(\mathbf{K}_{j}+\mathbf{k}).\mathbf{R}_{i}}a_{\mathbf{K}_{j}+\mathbf{k}}$ ($a_{i}^{\dag}=\frac{1}{\sqrt{N}}\sum_{\mathbf{k}}\sum_{j}e^{-i(\mathbf{K}_{j}+\mathbf{k}).\mathbf{R}_{i}}a^{\dag}_{\mathbf{K}_{j}+\mathbf{k}}$) where N is the number of unit cells of pristine graphene, and $\mathbf{k}$ runs over the first Brillouin of graphene superlattice, red hexagonal in \ref{Fig01}(b).
The Fourier transform of Eq. (\ref{eq01}) is then,
\begin{eqnarray}\label{eq02}
H(\mathbf{k})&=&\sum_{\mathbf{k}}\sum_{j}(\epsilon^{\ast}(\mathbf{K}_{j}+\mathbf{k})b_{\mathbf{K}_{j}+\mathbf{k}}^{\dag}a_{\mathbf{K}_{j}
+\mathbf{k}}+h.c.)+\frac{u}{3}\sum_{\mathbf{k}}\sum_{j,j^{'}}a_{\mathbf{K}_{j}+\mathbf{k}}^{\dag}a_{\mathbf{K}_{j^{'}}+\mathbf{k}} \nonumber \\
&+&\Delta\sum_{\mathbf{k}}\sum_{j,j^{'}}([1-2\cos(2(j-j^{'})\pi/3)]\epsilon^{\ast}(\mathbf{K}_{j}+\mathbf{k})b_{\mathbf{K}_{j}+
\mathbf{k}}^{\dag}a_{\mathbf{K}_{j^{'}}+\mathbf{k}}+h.c.),
\end{eqnarray}
where $\epsilon^{\ast}(\mathbf{K}_{j}+\mathbf{k})=-t_{0}\sum_{n=1}^{3}e^{-i(\mathbf{K}_{j}+\mathbf{k}).\mathbf{\delta}_{n}}$ and $\Delta=\frac{\delta t}{3t_{0}}$.

By defining $\hat{\psi}_{\mathbf{k}}=(a_{\mathbf{\Gamma}+\mathbf{k}},b_{\mathbf{\Gamma}+\mathbf{k}},a_{\mathbf{K}+\mathbf{k}},b_{\mathbf{K}+\mathbf{k}}
,a_{\mathbf{K}^{'}+\mathbf{k}},b_{\mathbf{K}^{'}+\mathbf{k}})^{T}$ we can rewrite the Hamiltonian as
\begin{eqnarray}\label{eq03}
H(\mathbf{k})=\hat{\psi}^{\dag}_{\mathbf{k}} \left(
                                             \begin{array}{cc}
                                              \hat{H}_{H}   &  \hat{T}^{\dag}    \\
                                               \hat{T}     &   \hat{H}_{L}
                                               \end{array}
                                            \right)
              \hat{\psi}_{\mathbf{k}} ,
\end{eqnarray}
where
\begin{eqnarray}\label{eq04}
\hat{H}_{H}=\left(
              \begin{array}{cc}
                                \frac{u}{3}                     &    (1-\Delta)\epsilon(\mathbf{\Gamma}+\mathbf{k})      \\
        (1-\Delta)\epsilon^{\ast}(\mathbf{\Gamma}+\mathbf{k})   &                       0
              \end{array}
                       \right),
\end{eqnarray}
and
\begin{eqnarray}\label{eq05}
\hat{H}_{L}=\left(
              \begin{array}{cccc}
                        \frac{u}{3}                      &       (1-\Delta)\epsilon(\mathbf{K}+\mathbf{k})       &
                        \frac{u}{3}                      &      2\Delta\epsilon(\mathbf{K}^{'}+\mathbf{k})           \\
   (1-\Delta)\epsilon^{\ast}(\mathbf{K}+\mathbf{k})      &                         0                             &
      2\Delta\epsilon^{\ast}(\mathbf{K}+\mathbf{k})      &                         0                                 \\
                        \frac{u}{3}                      &         2\Delta\epsilon(\mathbf{K}+\mathbf{k})        &
                        \frac{u}{3}                      &     (1-\Delta)\epsilon(\mathbf{K}^{'}+\mathbf{k})         \\
    2\Delta\epsilon^{\ast}(\mathbf{K}^{'}+\mathbf{k})    &                         0                             &
  (1-\Delta)\epsilon^{\ast}(\mathbf{K}^{'}+\mathbf{k})   &                         0
              \end{array}
                       \right),
\end{eqnarray}
are the high and low energy sectors of $\hat{H}(\mathbf{k})$, respectively, and
\begin{eqnarray}\label{eq06}
\hat{T}=\left(
              \begin{array}{cccc}
                        \frac{u}{3}                      &      2\Delta\epsilon(\mathbf{K}+\mathbf{k})      &
                        \frac{u}{3}                      &    2\Delta\epsilon(\mathbf{K}^{'}+\mathbf{k})        \\
   2\Delta\epsilon^{\ast}(\mathbf{\Gamma}+\mathbf{k})    &                         0                        &
   2\Delta\epsilon^{\ast}(\mathbf{\Gamma}+\mathbf{k})    &                         0
              \end{array}
                       \right),
\end{eqnarray}
is the coupling between them.
Then, we consider the Schrodinger equations applied to Eq. (\ref{eq03}),
\begin{eqnarray}\label{eq07}
\hat{H}_{H}\hat{\Psi}_{H}+\hat{T}\hat{\Psi}_{L}=E\hat{\Psi}_{H},
\end{eqnarray}
and
\begin{eqnarray}\label{eq08}
\hat{T}^{\dag}\hat{\Psi}_{H}+\hat{H}_{L}\hat{\Psi}_{L}=E\hat{\Psi}_{L},
\end{eqnarray}
on each space, where $\hat{\Psi}_{H}$ and $\hat{\Psi}_{L}$ are the components of the eigenfunction, $\hat{\Psi}=(\hat{\Psi}_{H},\hat{\Psi}_{L})$. One can obtain $\hat{\Psi}_{H}$ from Eq. (\ref{eq07}) and use it on Eq. (\ref{eq08}) to obtain
\begin{eqnarray}\label{eq09}
\hat{H}_{eff}=\hat{H}_{L}+\hat{T}^{\dag}(E\hat{1}-\hat{H}_{L})^{-1}\hat{T},
\end{eqnarray}
which is the effective Hamiltonian for the $\hat{\Psi}_{L}$ component. Then, we expand $\epsilon(\mathbf{K}_{j}+\mathbf{k})$ and their complex conjugates in Eqs. (\ref{eq04}-\ref{eq06}) up to the first order in $\mathbf{k}$, $\epsilon(\mathbf{\Gamma}+\mathbf{k})\approx-3t_{0}$, $\epsilon(\mathbf{K}+\mathbf{k})\approx\frac{3}{2}t_{0}a(k_{x}+ik_{y})$ and $\epsilon(\mathbf{K}^{'}+\mathbf{k})\approx-\frac{3}{2}t_{0}a(k_{x}-ik_{y})$, and insert them into Eq. (\ref{eq09}). Therefore, we arrive at
\begin{eqnarray}\label{eq10}
\hat{H}_{eff}=\left(
              \begin{array}{cccc}
             U             &       v_{B}(k_{x}+ik_{y})     &
             U             &      -v_{C}(k_{x}-ik_{y})         \\
    v_{B}(k_{x}-ik_{y})    &                 0             &
    v_{C}(k_{x}-ik_{y})    &                 0                 \\
             U             &      v_{C}(k_{x}+ik_{y})      &
             U             &     -v_{B}(k_{x}-ik_{y})          \\
  -v_{C}(k_{x}+ik_{y})     &                 0             &
  -v_{B}(k_{x}+ik_{y})     &                 0
              \end{array}
                       \right),
\end{eqnarray}
where $U=\frac{u}{3}(1-\frac{4\Delta}{1-\Delta}+\frac{4\Delta^{2}}{(1-\Delta)^{2}})$, $v_{B}=\frac{3}{2}t_{0}a(1-\Delta-\frac{4\Delta^{2}}{1-\Delta})$, and $v_{C}=\frac{3}{2}t_{0}a(2\Delta-\frac{4\Delta^{2}}{1-\Delta})$ (Notice that although Eq. (\ref{eq10}) is convenient to obtain the paramagnetic current operator, to get the diamagnetic part of the current operator we must keep the terms up to second order in $\mathbf{k}$). The energy bands of the low-energy effective Hamiltonian are
\begin{eqnarray}\label{eq11}
E_{1\mathbf{k}}&=&-E_{2\mathbf{k}}=-(v_{B}-v_{C})k, \nonumber \\
E_{3\mathbf{k}}&=&U-\sqrt{(v_{B}+v_{C})^{2}k^{2}+U^{2}}, \nonumber \\
E_{4\mathbf{k}}&=&U+\sqrt{(v_{B}+v_{C})^{2}k^{2}+U^{2}},
\end{eqnarray}
where $k=\sqrt{k^{2}_{x}+k^{2}_{y}}$. Figure \ref{Fig02} displays the exact band structure resulting from Eq. (\ref{eq01}) or (\ref{eq02}) in comparison with that obtained from our low-energy effective Hamiltonian for different values of $u$ and $\delta t$, (a) $u=0$ and $\delta t=0.1t_{0}$, (b) $u=0.1t_{0}$ and $\delta t=0$, (c) $u=0.1t_{0}$ and $\delta t=0.1t_{0}$. An excellent agreement between the energy bands of the effective Hamiltonian and the exact ones in the low-energy regime is clear. One can see that, the Y-shaped Kekul\'{e} distortion breaks the valley degeneracy and couples the energy bands at $\mathbf{K}$ and $\mathbf{K}^{'}$ in the firs brillouin zone, resulting in two concentric energy bands with different velocities at $\mathbf{\Gamma}$ point, but contrary to what has been reported in previous works \cite{Gamayun1,Andrade3}, their deviation from initial velocity is not symmetric. According to our results, different velocities are $(1-3\Delta)v_{F}$ and $(1+\Delta-\frac{8\Delta^{2}}{1-\Delta})v_{F}$ with $v_{F}=\frac{3}{2}t_{0}a$ being the initial Fermi velocity. On the other hand, the on-site potential deviation, in addition two breaking valley symmetry, has two other effects. First, it causes two energy bands gain effective mass. Second, it shifts the massive energy bands, resulting in a particle-hole symmetry breaking and lifting the four fold degeneracy at Dirac point. These effects lead to the appearance of attractive optical signatures, which are discussed in the next section.

\begin{figure}
\begin{center}
\includegraphics[width=17.5cm,angle=0]{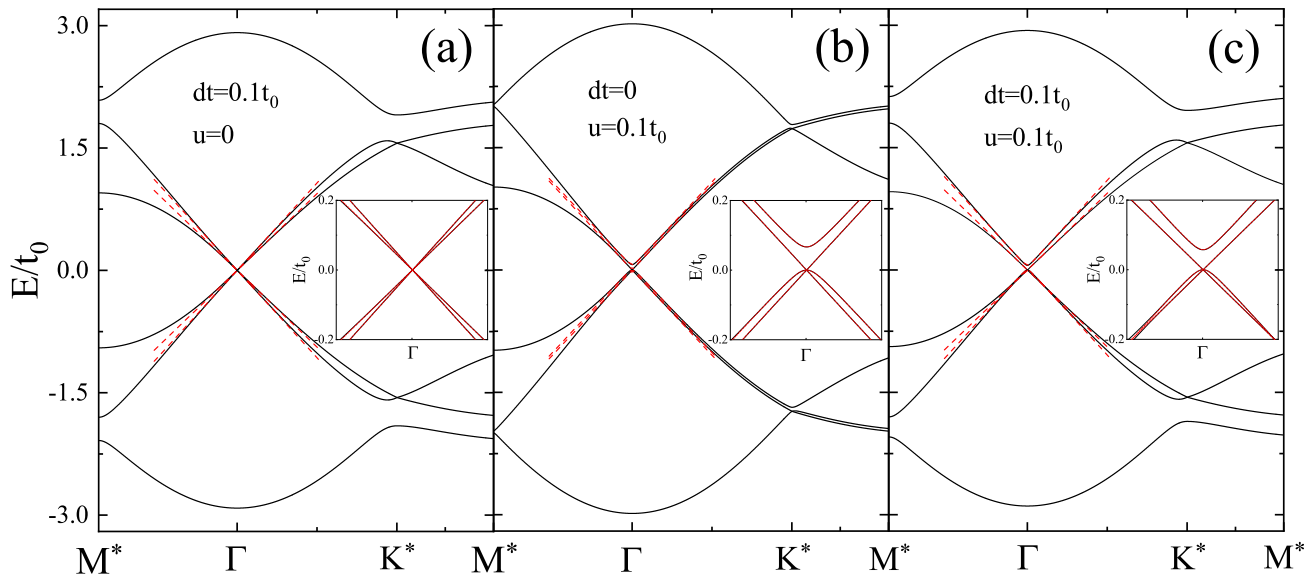}
\caption{Comparison of the exact energy bands resulting from Eq. (\ref{eq01}) or (\ref{eq02}) with the effective low-energy bands, Eq. (\ref{eq11}), for different values of the hopping and on-site energy deviations, (a) $\delta t=0.1t_{0}$ and $u=0$, (b) $\delta t=0$ and $u=0.1t_{0}$ and (c) $\delta t=0.1t_{0}$ and $u=0.1t_{0}$. $\mathbf{M^{\ast}}=(+2\pi/3\sqrt{3},0)$, $\mathbf{\Gamma}=(0,0)$ and $\mathbf{K^{\ast}}=(+2\pi/3\sqrt{3},+2\pi/9)$ are high-symmetry points in the first brillouin zone of graphene superlattice. See red hexagonal in Fig. \ref{Fig01}(b) .}
\label{Fig02}
\end{center}
\end{figure}

\subsection{Current operator and Kubo Formula}
\label{sec02-2}

The finite frequency conductivity calculated by Kubo Formula is given by\cite{Stauber1}
\begin{eqnarray}\label{eq12}
\sigma_{\alpha\alpha}(\Omega)=\frac{\Pi_{\alpha\alpha}(\Omega+i0^{+})}{i\hbar(\Omega+i0^{+})S}+\frac{<j_{\alpha}^{D}>}{i(\Omega+i0^{+})S},
\end{eqnarray}
where $S$ is the area of the sample, $\alpha$ denotes $x$ or $y$, $\Omega$ is the frequency, $\Pi_{\alpha\alpha}(\Omega+i0^{+})$ is the correlation function of the paramagnetic current operator and $<j_{\alpha}^{D}>$ is the expectation value of the diamagnetic part of current operator. The function $\Pi_{\alpha\alpha}(\Omega+i0^{+})$ is obtained by analytical continuum of the corresponding Matsubara current-current correlation function defined as\cite{Mahan1,Nicol1,Mohammadi3}
\begin{eqnarray}\label{eq13}
\Pi_{\alpha\alpha}(i\omega_{n})=\int_{0}^{\hbar\beta}d\tau e^{i\omega_{n}\tau}<T_{\tau}j_{\alpha}^{P}(\tau)j_{\alpha}^{P}(0)>,
\end{eqnarray}
where $i\omega_{n}$ are the bosonic Matsubara frequencies, $\beta=1/k_{B}T$ with $k_{B}$ the Boltzmann constant and T the temperature, and $T_{\tau}$ is time ordering operator. One can show that the paramagnetic current operator can be written as \cite{Nicol1} $j_{\alpha}^{P}=-g_{s}e\Sigma_{\mathbf{k}}\hat{\psi}^{\dag}_{\mathbf{k},L}\hat{v}_{\alpha}\hat{\psi}_{\mathbf{k},L}$ where $g_{s}=2$ is the spin degeneracy, $e$ is the electron charge, $\hat{\psi}_{\mathbf{k},L}$ is the low-energy component of $\hat{\psi}_{\mathbf{k}}=(\hat{\psi}_{\mathbf{k},H},\hat{\psi}_{\mathbf{k},L})^{T}$, and $\hat{v}_{\alpha}=\partial \hat{H}_{eff}/\hbar\partial k_{\alpha}$. Consequently, the current-current correlation function can then be written in the usual bubble approximation as\cite{Mahan1,Nicol1}
\begin{eqnarray}\label{eq14}
\Pi_{\alpha\alpha}(i\nu_{m})=\frac{g_{s}e^{2}}{\beta\hbar}\sum_{i\nu_{m}}\sum_{\mathbf{k}} Tr[\hat{v}_{\alpha}\hat{G}(i\nu_{m}+i\omega_{n},\mathbf{k})\hat{v}_{\alpha}\hat{G}(i\nu_{m},\mathbf{k})],
\end{eqnarray}
where $\hat{G}(i\nu_{m},\mathbf{k})=(i\nu_{m}\hat{1}-\hat{H}_{eff}/\hbar)^{-1}$, $Tr[\hat{A}]$ denotes the trace of $\hat{A}$ matrix, and $i\nu_{m}$ are the fermionic Matsubara frequencies. Using the spectral representation of the Green's function matrix,\cite{Nicol1,Mohammadi4}
\begin{eqnarray}\label{eq15}
\hat{G}(z,\mathbf{k})=\int_{-\infty}^{+\infty}\frac{d\omega^{'}}{2\pi}\frac{\hat{A}(\omega^{'},\mathbf{k})}{z-\omega^{'}},
\end{eqnarray}
the imaginary and real part of the current-current correlation function can be written as
\begin{eqnarray}\label{eq16}
\Im\Pi_{xx}(\Omega+i0^{+})=-\frac{\pi g_{s}e^{2}}{\hbar^{2}}\sum_{\mathbf{k}}\sum_{i,j}\chi(E_{j\mathbf{k}},E_{i\mathbf{k}})
[n_{F}(E_{j\mathbf{k}})-n_{F}(E_{i\mathbf{k}})]\delta(\Omega+E_{j\mathbf{k}}/\hbar-E_{i\mathbf{k}}/\hbar),
\end{eqnarray}
and
\begin{eqnarray}\label{eq17}
\Re\Pi_{xx}(\Omega+i0^{+})=\frac{g_{s}e^{2}}{\hbar^{2}}\sum_{\mathbf{k}}\sum_{i,j}\chi(E_{i\mathbf{k}},E_{j\mathbf{k}})
\frac{n_{F}(E_{j\mathbf{k}})-n_{F}(E_{i\mathbf{k}})}{\Omega+E_{j\mathbf{k}}/\hbar-E_{i\mathbf{k}}/\hbar},
\end{eqnarray}
where $n_{F}(x)=1/(1+exp[(x-\mu)/k_{B}T])$ is Fermi-Dirac distribution function and
\begin{eqnarray}\label{eq18}
\chi(E_{i\mathbf{k}},E_{j\mathbf{k}})&=&4(v^{2}_{B}\cos(2\varphi_{\mathbf{k}})+v^{2}_{C})\mathcal{M}_{12,E_{i\mathbf{k}}}\mathcal{M}_{12,
E_{j\mathbf{k}}}+4(v^{2}_{B}+v^{2}_{C}\cos(2\varphi_{\mathbf{k}}))\mathcal{M}_{12,E_{i\mathbf{k}}}\mathcal{M}_{12,E_{j\mathbf{k}}} \nonumber \\
&+&4v_{B}v_{C}(1+\cos(2\varphi_{\mathbf{k}}))[\mathcal{M}_{12,E_{i\mathbf{k}}}\mathcal{M}_{13,E_{j\mathbf{k}}}
+\mathcal{M}_{13,E_{i\mathbf{k}}}\mathcal{M}_{12,E_{j\mathbf{k}}}] \nonumber \\
&+&2(v^{2}_{B}+v^{2}_{C})[\mathcal{M}_{11,E_{i\mathbf{k}}}\mathcal{M}_{22,E_{j\mathbf{k}}}
+\mathcal{M}_{22,E_{i\mathbf{k}}}\mathcal{M}_{11,E_{j\mathbf{k}}}] \nonumber \\
&+&4v_{B}v_{C}[\mathcal{M}_{11,E_{i\mathbf{k}}}\mathcal{M}_{23,E_{j\mathbf{k}}}
+\mathcal{M}_{23,E_{i\mathbf{k}}}\mathcal{M}_{11,E_{j\mathbf{k}}}] \nonumber \\
&-&4v_{B}v_{C}\cos(2\varphi_{\mathbf{k}})[\mathcal{M}_{14,E_{i\mathbf{k}}}\mathcal{M}_{22,E_{j\mathbf{k}}}
+\mathcal{M}_{22,E_{i\mathbf{k}}}\mathcal{M}_{14,E_{j\mathbf{k}}}] \nonumber \\
&-&2(v^{2}_{B}+v^{2}_{C})[\mathcal{M}_{14,E_{i\mathbf{k}}}\mathcal{M}_{23,E_{j\mathbf{k}}}
+\mathcal{M}_{23,E_{i\mathbf{k}}}\mathcal{M}_{14,E_{j\mathbf{k}}}],
\end{eqnarray}
in which $\varphi_{\mathbf{k}}=\tan^{-1}(k_{y}/k_{x})$ and $\mathcal{M}_{mn,E_{i\mathbf{k}}}$ are the coefficients of delta functions $\delta(\omega-E_{i\mathbf{k}}/\hbar)$ in the matrix components of the spectral function, $A_{mn}(\omega,\mathbf{k})=2\pi\sum_{i}\mathcal{M}_{mn,E_{i\mathbf{k}}}\delta(\omega-E_{i\mathbf{k}}/\hbar)$.

To obtain $<j_{\alpha}^{D}>$ we keep the terms up to second order in $\mathbf{k}$ in the expansion of $\epsilon(\mathbf{K}_{j}+\mathbf{k})$ and its complex conjugate in Eq. (\ref{eq09}), so we can write the diamagnetic current operator as\cite{Peres1} $j^{D}_{\alpha}=-g_{s}e^{2}\Sigma_{\mathbf{k}}\hat{\psi}^{\dag}_{\mathbf{k},L}\hat{w}_{\alpha}\hat{\psi}_{\mathbf{k},L}$,
where
\begin{eqnarray}\label{eq19}
\hat{w}_{x}=\frac{1}{\hbar^{2}}\frac{\partial^{2}\hat{H}_{eff}}{\partial k^{2}_{x}}=\frac{a}{2\hbar^{2}}\left(
              \begin{array}{cccc}
        0      &   -v_{B}    &      0      &    v_{C}     \\
     -v_{B}    &      0      &   -v_{C}    &      0       \\
        0      &   -v_{C}    &      0      &    v_{B}     \\
      v_{C}    &      0      &    v_{B}    &      0
              \end{array}
                       \right).
\end{eqnarray}
After inserting above equations into $<j_{\alpha}^{D}>$ and performing some straightforward calculations\cite{Bruus1}, we reach the following statement
\begin{eqnarray}\label{eq20}
<j_{x}^{D}>=\frac{g_{s}ae^{2}}{2\hbar^{^{2}}}\sum_{\mathbf{k}}\sum_{i} \Lambda(E_{i\mathbf{k}})n_{F}(E_{i\mathbf{k}}),
\end{eqnarray}
where
\begin{eqnarray}\label{eq21}
\Lambda(E_{i\mathbf{k}})=4v_{B}\mathcal{M}_{12,E_{i\mathbf{k}}}\cos(2\varphi_{\mathbf{k}})+2v_{C}\mathcal{M}_{14,E_{i\mathbf{k}}}\cos(4\varphi_{\mathbf{k}})
+2v_{C}\mathcal{M}_{23,E_{i\mathbf{k}}}.
\end{eqnarray}

The real part of the finite frequency conductivity, which is related to the optical quantities such optical absorption and reflectivity, is given by
\begin{eqnarray}\label{eq22}
\Re\sigma_{\alpha\alpha}(\Omega)=D\delta(\Omega)+\frac{\Im\Pi_{\alpha\alpha}(\Omega+i0^{+})}{\hbar\Omega S},
\end{eqnarray}
with
\begin{eqnarray}\label{eq23}
D=-\pi\frac{<j_{\alpha}^{D}>}{S}-\pi\frac{\Re\Pi_{\alpha\alpha}(\Omega+i0^{+})}{\hbar S}.
\end{eqnarray}
called Drude weight. Finally, by inserting Eqs. (\ref{eq16}), (\ref{eq17}) and (\ref{eq20}) into (\ref{eq22}), we arrive at the following equation for the real part of the optical conductivity
\begin{eqnarray}\label{eq24}
\Re\sigma_{\alpha\alpha}(\Omega)/\sigma_{0}=&-&\frac{2\pi ag_{s}}{\hbar}\frac{1}{S}\sum_{\mathbf{k}}\sum_{i} \Lambda(E_{i\mathbf{k}})n_{F}(E_{i\mathbf{k}})\delta(\Omega) \nonumber \\
&-&\frac{4\pi g_{s}}{\hbar^{2}}\frac{1}{S}\sum_{\mathbf{k}}\sum_{i,j}\chi(E_{i\mathbf{k}},E_{j\mathbf{k}})
\frac{n_{F}(E_{j\mathbf{k}})-n_{F}(E_{i\mathbf{k}})}{\Omega+E_{j\mathbf{k}}/\hbar-E_{i\mathbf{k}}/\hbar}\delta(\Omega) \\
&-&\frac{4\pi g_{s}}{\hbar^{2}}\frac{1}{S}\sum_{\mathbf{k}}\sum_{i,j}\chi(E_{i\mathbf{k}},E_{j\mathbf{k}})
\frac{[n_{F}(E_{j\mathbf{k}})-n_{F}(E_{i\mathbf{k}})]}{\Omega}\delta(\Omega+E_{j\mathbf{k}}/\hbar-E_{i\mathbf{k}}/\hbar) \nonumber,
\end{eqnarray}
where $\sigma_{0}=\frac{e^{2}}{4\hbar}$ is the conductivity of the pristine graphene.

Eqs. (\ref{eq18}), (\ref{eq21}), and (\ref{eq24}) along with the low-energy effective Hamiltonian and its energy bands are our main results, which are solved numerically in the next section to investigate the effects of the kekul\'{e} distortion on the optical conductivity of graphene at low energy regime.

\section{Results and discussion}
\label{sec03}

Here we present our results exploring the effects of Y-shaped Kekul\'{e} distortion on the optical conductivity of graphene, which is obtained by evaluating Eq. (\ref{eq24}) numerically. Although it is expected that the Y-shaped Kekul\'{e} distortion leads to a deviation in both hopping and on-site energies, but in order to clarify the role of each of them separately, in addition to the $\delta t\neq0$ and $u\neq0$ case which is expected to be more realistic, we also investigate the cases in which $\delta t\neq0$ or $u\neq0$. For the numerical evaluation of Eq. (\ref{eq24}), we use the Lorentzian representation of the delta function, $\delta(x)=(\eta/\pi)/(x^{2}+\eta^{2})$ with broadening $\eta$ reflecting the effect of electron scattering from disorder, $1/\tau_{imp}=2\eta$, phenomenologically. In all calculations we set $\eta=0.00002t_{0}$. Furthermore, we put $t_{0}$ as the energy scale and the optical conductivity curves are scaled by $\sigma_{0}$. So, $\sigma_{xx}/\sigma_{0}$ is calculated and plotted as a function of $\hbar\Omega/t_{0}$ with $\hbar\Omega$ being the photon energy.

\textit{Zero chemical potential}: Our results for the optical conductivity at zero chemical potential and temperature are shown in Fig. \ref{Fig03}, in which the hopping and on-site energy deviations are (a) $\delta t=0.10t_{0}$ and $u=0$, (b) $\delta t=0$ and $u=0.10t_{0}$ and (c) $\delta t=0.10t_{0}$ and $u=0.10t_{0}$. The results for  Kekul\'{e}-patterned (pristine) graphene are shown by solid magenta (dashed black) curves. All possible optical transitions are displayed in the lower insets of Fig. \ref{Fig03} and the resulting structures in the optical conductivity are identified by arrows with the same color on the curves. The upper insets exhibit the contribution of each transition in optical conductivity separately identified by the curve with same color. In the case of $\delta t=0.10t_{0}$ and $u=0$, the optical conductivity displays a flat interband response at all frequencies which also is found in undoped pristine graphene\cite{Novoselov1,Peres2}, reflecting the linear band structure of Kekul\'{e}-patterned graphene near $\mathbf{\Gamma}$ point. This is to be contrast with the case of $u\neq0$ shown in the middle panels Fig. \ref{Fig03}, in which a dip-peak structure is also seen around $2U$. The appearance of this structure in the optical conductivity, which can be detected as a optical signature for the Y-shaped Kekul\'{e} distortion in single layer graphene at the charge neutrality, is explained by examining the contribution of each of the transitions in the optical conductivity. In the presence of the on-site energy deviation $E_{3\mathbf{k}}\rightarrow E_{2\mathbf{k}}$ transition is possible for all $\hbar\Omega>0$, but $E_{1\mathbf{k}}\rightarrow E_{4\mathbf{k}}$ transition, due to the presence of the energy gap, is restricted to $\hbar\Omega>2U$. The $E_{3\mathbf{k}}\rightarrow E_{2\mathbf{k}}$ ($E_{1\mathbf{k}}\rightarrow E_{4\mathbf{k}}$) transition, as shown in the upper inset of the middle frame of Fig. \ref{Fig03}, starts at zero ($2U$) photon energy with a spectral weight that decreases (increases) by increasing the photon energy and finally approaches $\sigma_{0}/2$ at high photon energies. These facts manifest themselves as a dip in the curve of the total optical conductivity. This spectral weight lost, for $\hbar\Omega>2U$, is compensate by the appearance of a new transition which starts at $\hbar\Omega=2U$, $E_{3\mathbf{k}}\rightarrow E_{4\mathbf{k}}$, and its large spectral weight also leads to the appearance of the peak structure. At high photon energies, where the effect of the energy gap becomes insignificant, $E_{2\mathbf{k}}$ and $E_{4\mathbf{k}}$ bands inherit the chirality of the energy bands in $\mathbf{K}$ and $\mathbf{K}^{'}$ valleys in pristine graphene and become completely chiral and anti-chiral respectively. So, the optical transitions between them becomes forbidden\cite{Herrera3}. Therefore, at high photon energies only $E_{1\mathbf{k}}\rightarrow E_{4\mathbf{k}}$ and $E_{2\mathbf{k}}\rightarrow E_{4\mathbf{k}}$ transitions contribute to the optical conductivity and the usual flat interband response is recovered. Returning to the case of $\delta t=0.10t_{0}$ and $u=0.10t_{0}$ (Fig. \ref{Fig03}(c)), one can see that the only additional change caused by the simultaneous presence of the hopping and the on-site energy deviations is that the dip-peak structure of the optical conductivity is shifted to lower energies originating from the dependence of $U$ on the hopping energy deviation, $U=\frac{u}{3}(1-\frac{4\Delta}{1-\Delta}+\frac{4\Delta^{2}}{(1-\Delta)^{2}})$.

\begin{figure}
\begin{center}
\includegraphics[width=17.5cm,angle=0]{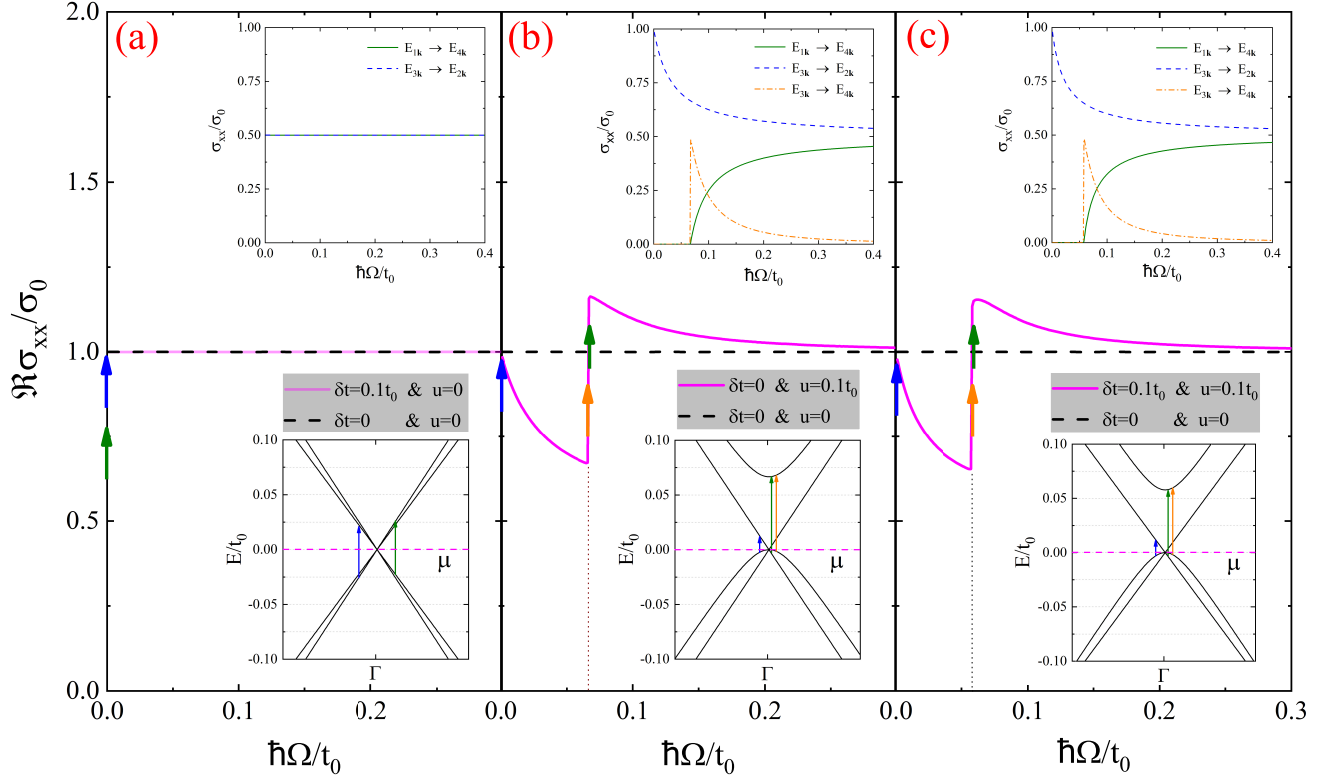}
\caption{The normalized optical conductivity, $\sigma_{xx}/\sigma_{0}$, of undoped Kekul\'{e}-patterned graphene (solid magenta curves) as function of $\hbar\Omega/t_{0}$ at zero temperature for (a) $\delta t=0.1t_{0}$ and $u=0$, (b) $\delta t=0$ and $u=0.1t_{0}$, and (c) $\delta t=0.1t_{0}$ and $u=0.1t_{0}$, compared with that of pristine graphene (dashed black curves). The lower insets show the band structures around $\mathbf{\Gamma}$ point for each case, and allowed optical transitions indicated by arrows which give rise to structure in the conductivity curves shown by arrows with same color on plots. The upper insets in each panel displays the contribution of each of the optical transitions in the optical conductivity separately.}
\label{Fig03}
\end{center}
\end{figure}

\textit{Finite chemical potential}: Figure \ref{Fig04} shows our numerical results for the optical conductivity of Y-shaped Kekul\'{e}-patterned graphene in comparison with that of pristine graphene for $\mu>0$. The upper (lower) panels display the result for $\mu=+0.05t_{0}$ ($\mu=+0.10t_{0}$), in which the left, middle and panels is for $\delta t=0.10t_{0}~~u=0$, $\delta t=0.0~~u=0.10t_{0}$ and $\delta t=0.10t_{0}~~u=0.10t_{0}$ respectively. As in Fig. \ref{Fig03}, the solid magenta (dashed black) curves display the optical conductivity of Kekul\'{e}-patterned (pristine) graphene, and in each panel, the inset shows the corresponding band structure and the allowed inter-band optical transitions. The colored arrows in the band structure show the allowed optical transitions which give rise to some structures in the optical conductivity identified by arrows with same color on the plots. Figure \ref{Fig05} shows the same plots as Fig. \ref{Fig04}, but for other different values of the chemical potentials, $\mu=-0.05t_{0}$ (upper panels) and $\mu=-0.10t_{0}$ (lower panels). For $\mu\neq0$, as can be seen in all panels of Figs. \ref{Fig04} and \ref{Fig05}, there is the usual Drude conductivity arising from the intra-band transitions. Starting from the case of $\delta t=0.10t_{0}$ and $u=0$,  shown in Figs. \ref{Fig04}(a) and \ref{Fig03}(d), one can see that the interband absorption edge that is at $2\mu$ for pristine graphene\cite{Peres1} is now splitting into two edges moving to lower and higher photon energies. So, the optical conductivity exhibits a two-steps absorption which onset at $2\mu\frac{v_{B}}{v_{B}+v_{C}}$ and $2\mu\frac{v_{B}}{v_{B}-v_{C}}$ photon energies coming from the interband transitions between different valleys, $E_{1\mathbf{k}}\rightarrow E_{4\mathbf{k}}$ and $E_{3\mathbf{k}}\rightarrow E_{2\mathbf{k}}$. The presence of the Kekule distortion, which couples different valleys, also leads to the appearance of a sharp peak in the optical conductivity which occurs in the range of $2\mu\frac{v_{C}}{v_{B}+v_{C}}$ to $2\mu\frac{v_{C}}{v_{B}-v_{C}}$ and is due to the inter-valley transitions in the conduction band, shown by red arrow in the insets. See also (a), (d), (g) and (j) panels of Fig. \ref{Fig06} in which the contribution of all transitions of Figs. \ref{Fig04} and \ref{Fig05} has been displayed. From Fig. \ref{Fig04} (d) one can also see that increasing the chemical potential enhances the splitting of the interband absorption edges and moves them further to higher and lower energies. It also shifts the sharp peak to higher energies and increase its spectral weight.

The effects of nonzero on-site energy deviation on the optical conductivity has been addressed in (b) and (e) panels of Fig. \ref{Fig04}. One can see that due to the gapped nature of $E_{4\mathbf{k}}$ band, which increases the energy required for $E_{3\mathbf{k}}\rightarrow E_{4\mathbf{k}}$ transition, the sharp peak in optical conductivity is enhanced and moved to higher energies, with a nonzero spectral weight in the energy range $U+\sqrt{U^{2}+[(v_{B}+v_{C})/(v_{B}-v_{C})]^{2}\mu^{2}}<\hbar\Omega<2U$. If the chemical potential increases and cuts $E_{4\mathbf{k}}$ band, due to the occurrence of a phenomenon called \textit{band nesting resonance}\cite{Carvalho1,Rashidian1,Mennel1}, the spectral weight of the peak, as can be seen in Fig. \ref{Fig04}(b), increases sharply and its broadening becomes limited to a small range of photon energies, facilitating its observation as an optical signature for the Y-shaped Kekul\'{e} distortion with nonzero on-site energy deviation.

In order to understand the origin of band nesting resonance, we rewrite the contribution of $E_{i\mathbf{k}}\rightarrow E_{j\mathbf{k}}$ transition in the optical conductivity in Eq. \ref{eq24}, $-\frac{4\pi g_{s}}{\hbar^{2}}\frac{1}{S}\sum_{\mathbf{k}}\chi(E_{j\mathbf{k}},E_{i\mathbf{k}})
\frac{[n_{F}(E_{i\mathbf{k}})-n_{F}(E_{j\mathbf{k}})]}{\Omega}\delta(\Omega+E_{i\mathbf{k}}/\hbar-E_{j\mathbf{k}}/\hbar)$, in a more illustrative form. If we consider cuts $S(E)$ of constant energy $E$, $E=E_{j\mathbf{k}}-E_{i\mathbf{k}}$, in the band structure, we can write $d^{2}\mathbf{k}=dS[d(E_{j\mathbf{k}}-E_{i\mathbf{k}})/|\nabla_{\mathbf{k}}(E_{j\mathbf{k}}-E_{i\mathbf{k}})|]$. So, we arrive at
\begin{eqnarray}\label{eq25}
\Re\sigma_{xx}^{E_{i}\rightarrow E_{j}}(\Omega)/\sigma_{0}=-\frac{g_{s}}{\pi\hbar\Omega}\int_{S(E)}
\frac{\chi(E_{j\mathbf{k}},E_{i\mathbf{k}})}{|\nabla_{\mathbf{k}}(E_{j\mathbf{k}}-E_{i\mathbf{k}})|}dS,
\end{eqnarray}
for the contribution of $E_{i\mathbf{k}}\rightarrow E_{j\mathbf{k}}$ transition in optical conductivity at \textit{zero temperature} indicating that strong peaks in the optical conductivity will come from regions in the spectrum where $|\nabla_{\mathbf{k}}(E_{j\mathbf{k}}-E_{i\mathbf{k}})|\approx 0$. One of the ways to fulfill the condition $|\nabla_{\mathbf{k}}(E_{j\mathbf{k}}-E_{i\mathbf{k}})|\approx 0$ is that the condition $|\nabla_{\mathbf{k}}E_{i\mathbf{k}}|\approx |\nabla_{\mathbf{k}}E_{j\mathbf{k}}|>0$, which occurs when two energy bands are approximately equispaced over regions in the Brillouin zone and called \textit{band nesting}, is satisfied. Band nesting results in a strong peak in the optical conductivity which is called band nesting resonance. One can easily check that the band nesting condition is satisfied for $E_{2\mathbf{k}}$ and $E_{4\mathbf{k}}$ bands in the case of $u\neq0$ when the chemical potential become larger than $2U$, as can be seen in Fig. \ref{Fig04}(e), resulting in the appearance of band nesting resonance in case of $\delta t=0$ and $u=0.1t_{0}$. Notice that the band nesting resonance also takes place for $E_{1\mathbf{k}}\rightarrow E_{3\mathbf{k}}$ transition but with a spectral weight more dependent on the chemical potential and restricted to limited ranges of the chemical potential.

The simultaneous presence of the hopping and on-site energy deviations makes the band nesting condition better fulfilled. Therefore, as Fig. \ref{Fig04}(f) shows, for the case of $\delta t=0.1t_{0}$ and $u=0.1t_{0}$ the intensity of the peak increases. Furthermore, since the nonzero hopping and on-site energy deviations decreases $U$ and shifts $E_{4\mathbf{k}}$ band to lower energies, the band nesting condition is also satisfied for a wide range of $\mathbf{k}$ in $E_{2\mathbf{k}}\rightarrow E_{4\mathbf{k}}$ transition and as a result, we see a large increase in the intensity of the peak in Fig. \ref{Fig04}(c) compared to that in Fig. \ref{Fig04}(b).

\begin{figure}
\begin{center}
\includegraphics[width=17.5cm,angle=0]{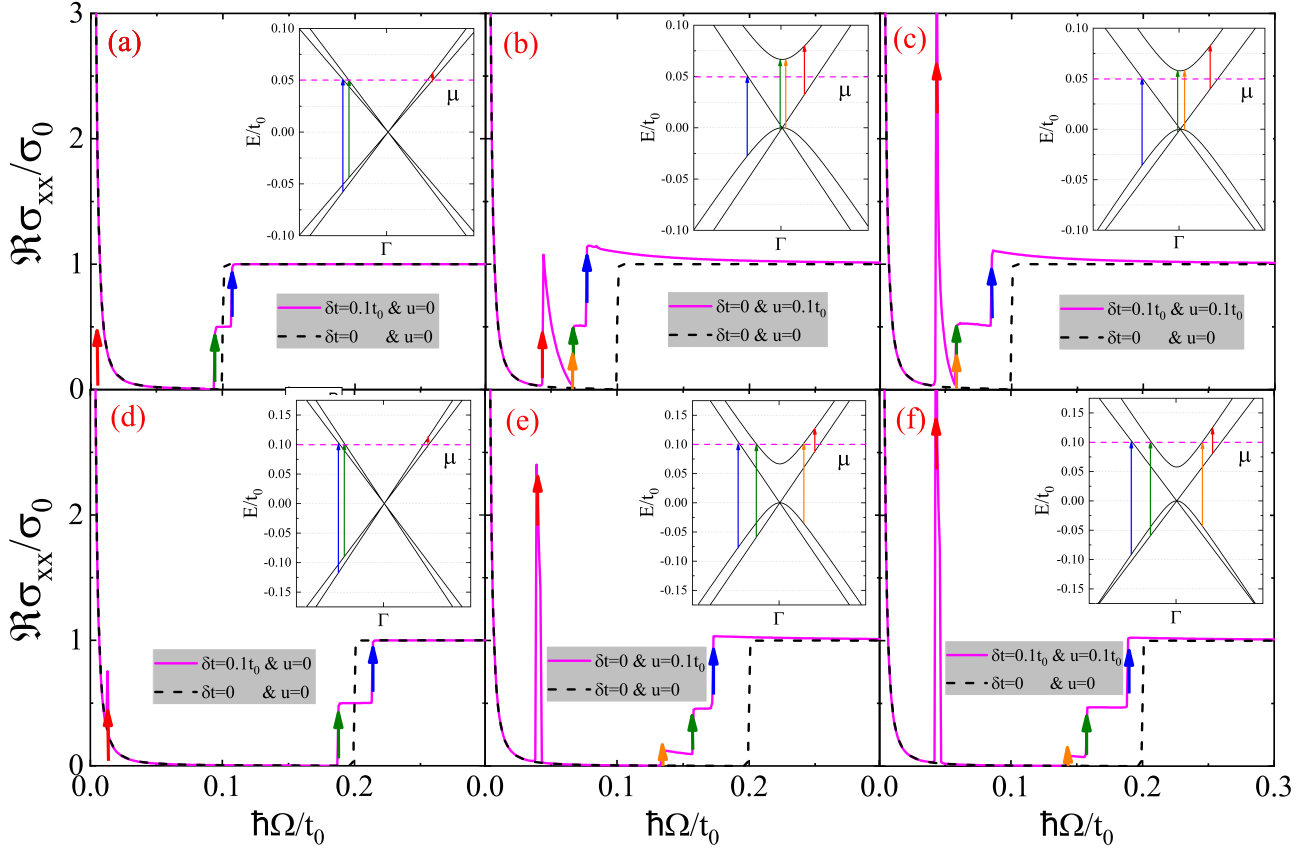}
\caption{The normalized optical conductivity, $\sigma_{xx}/\sigma_{0}$, of undoped Kekul\'{e}-patterned graphene (solid magenta curves) as function of $\hbar\Omega/t_{0}$ at zero temperature for different values of the chemical potential, $\mu=0.05t_{0}$ (upper panels) and $\mu=0.10t_{0}$ (lower panels), and different values of the on-site and hopping energy deviations, $\delta t=0.1t_{0}$ and $u=0$ (left panels), $\delta t=0$ and $u=0.1t_{0}$ (middle panels) and  $\delta t=0.1t_{0}$ and $u=0.1t_{0}$ (right panels), compared with that of pristine graphehe. The insets show the corresponding band structures around $\mathbf{\Gamma}$ point. The colored arrows in the band structure show the allowed optical transitions which give rise to some structures in the optical conductivity curves identified by arrows with same color on the plots.}
\label{Fig04}
\end{center}
\end{figure}

Returning to the cases of $u\neq0$ and $\mu=0.05t_{0}<2U$, shown in Figs. \ref{Fig04}(b) and \ref{Fig04}(c), one can see that the optical conductivity displays a similar two-steps interband absorbtion but with a interband absorbtion edges that, with respect to that in case of $\delta t=0.1t_{0}$ and $u=0$, moves to lower energies and onsets at $2U$ and $\mu+\sqrt{U^{2}+[(v_{B}+v_{C})/(v_{B}-v_{C})]^{2}\mu^{2}}-U$. A nonzero hopping energy deviation, as shown in Fig. \ref{Fig04}(c), enhances the splitting of the interband absorbtion edge. Moreover, the shape of the steps, in cases of $u\neq0$ and $\mu<2U$, which is typical of a gapped band structure \cite{Tabert1,Gusynin1} shows a peak at the absorbtion edge arising from a discontinuity in the electronic density of states. As the photon energy increases the usual flat interband response of pristine single layer graphene is recovered. By increasing the chemical potential, as can be seen in Figs. \ref{Fig04}(e) and \ref{Fig04}(f), $E_{2\mathbf{k}}\rightarrow E_{4\mathbf{k}}$ transitions become similar to that between linear energy bands leading to disappearance of the peak at the absorbtion edge. One can also see that in this case the optical conductivity displays a three-steps absorbtion arising from the splitting of the absorbtion edge of $E_{1\mathbf{k}}\rightarrow E_{4\mathbf{k}}$ and $E_{3\mathbf{k}}\rightarrow E_{4\mathbf{k}}$ transitions.

Figure \ref{Fig05} shows the same plots as Fig. \ref{Fig04}, but for $\mu=-0.05t_{0}$ (upper panels) and $\mu=-0.10t_{0}$ (lower panels). As expected, due to the presence of particle-hole symmetry, the optical conductivity in the case of $u=0$ does not change with changing the sign of the chemical potential, leading to similar plots as Figs. \ref{Fig04}(a) and \ref{Fig04}(d). This is in contrast to the case of nonzero on-site energy deviation, middle and right panels of Fig. \ref{Fig05}, in which the particle-hole symmetry is absent and the allowed optical transition due to the gapped nature of $E_{4\mathbf{k}}$ is shifted to higher energies. In these cases the optical conductivity, similar to that in the middle and right panels of Fig. \ref{Fig04}, displays a three-steps interband absorbtion but with the edges that move to higher energies. For $\mu<0$ the band nesting condition is satisfied for the case of $\delta t=0$ and $u=0.1t_{0}$ leading to appearance of band nesting resonance, but when both hopping and on-site energy deviation are nonzero the energy difference between $E_{1\mathbf{k}}$ and $E_{3\mathbf{k}}$ which determine the location and intensity of the intra-valley transition in the valance band (that turns into in band nesting resonance in the appropriate conditions\cite{Carvalho1}) is strongly dependent on the chemical potentials.

\begin{figure}
\begin{center}
\includegraphics[width=17.5cm,angle=0]{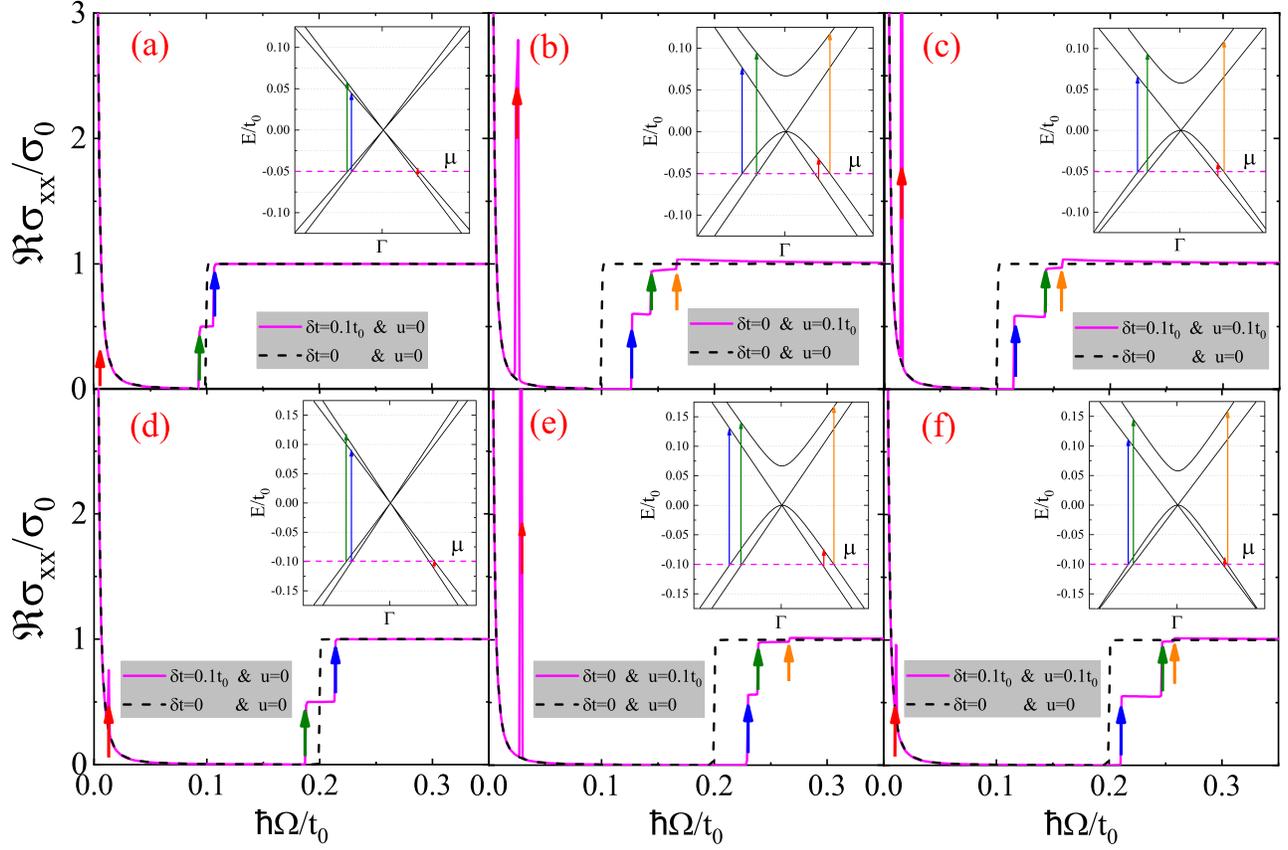}
\caption{Same plots as Fig. \ref{Fig04}, but for $\mu=-0.05t_{0}$ (upper panels) and $\mu=-0.10t_{0}$ (lower panels).}
\label{Fig05}
\end{center}
\end{figure}

\begin{figure}
\begin{center}
\includegraphics[width=17.5cm,angle=0]{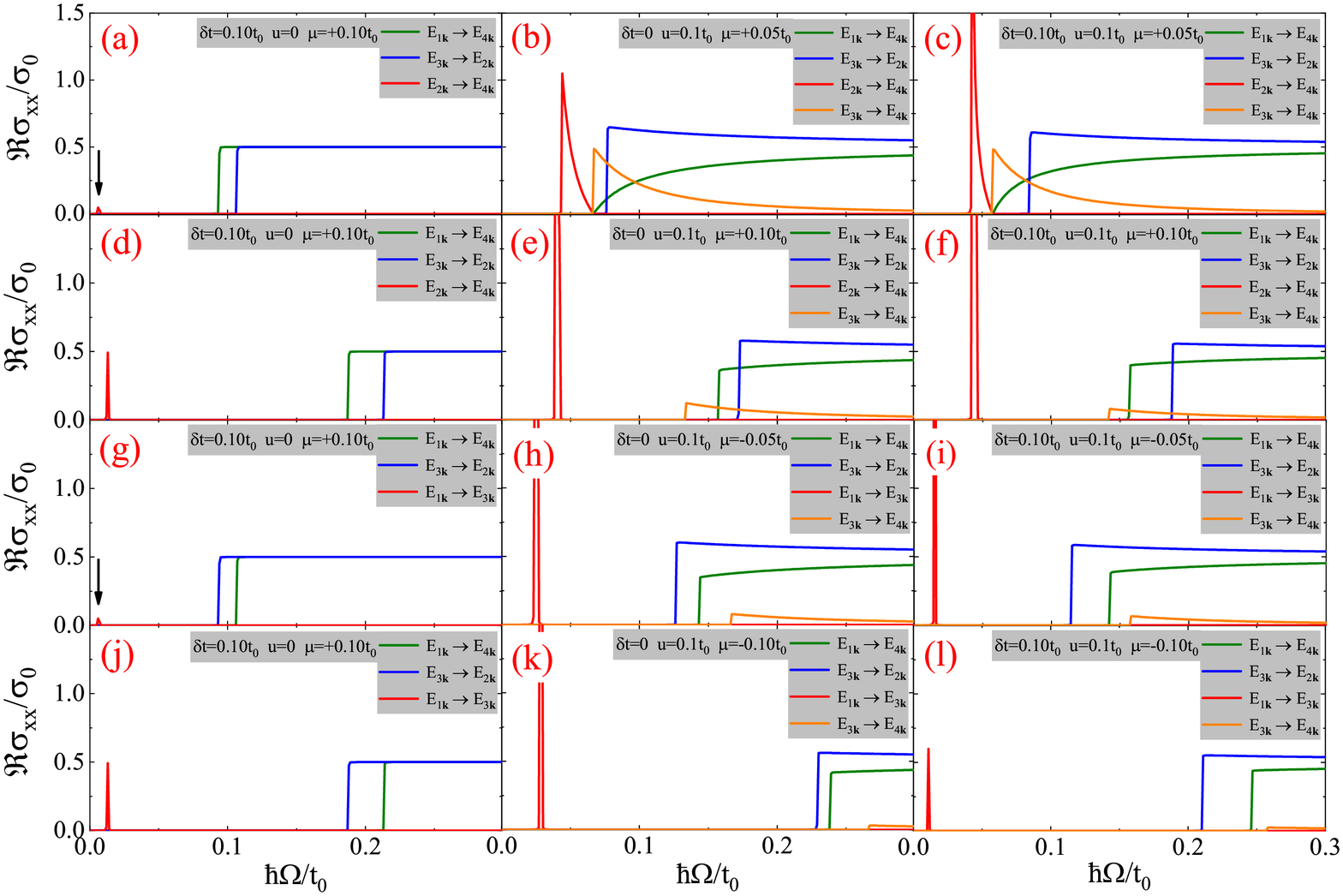}
\caption{Calculated normalized optical conductivity, $\sigma_{xx}/\sigma_{0}$, as function of $\hbar\Omega/t_{0}$ for each of the optical transitions shown in the insets of Figs. \ref{Fig04} and \ref{Fig04} separately.}
\label{Fig06}
\end{center}
\end{figure}

The main optical feature of the case of $\delta t\neq0$ and $u\neq0$, which is expected to be more realistic, is the occurrence of band nesting resonance. To show that the band nesting resonance is robust with respect to increasing temperature, the optical conductivity of Y-shaped Kekul-patterned graphene at different temperatures, $T=0$ (solid magenta curves) and $T=300~K$ (dashed black curves), has been displayed in Figure 7. One can see that increasing temperature only broadens the interband absorbtion edge. It is also clear that the band nesting resonance which occurs in both electron- and hole-doped cases is resistant against increasing temperature except when the chemical potential approaches the intersection of $E_{1\mathbf{k}}$ and $E_{3\mathbf{k}}$ energy bands ((d) panel). Therefore the occurrence of band nesting resonance is a significant optical signature to detect the Y-shaped Kekul\'{e} distortion in single layer graphene.

\begin{figure}
\begin{center}
\includegraphics[width=17.5cm,angle=0]{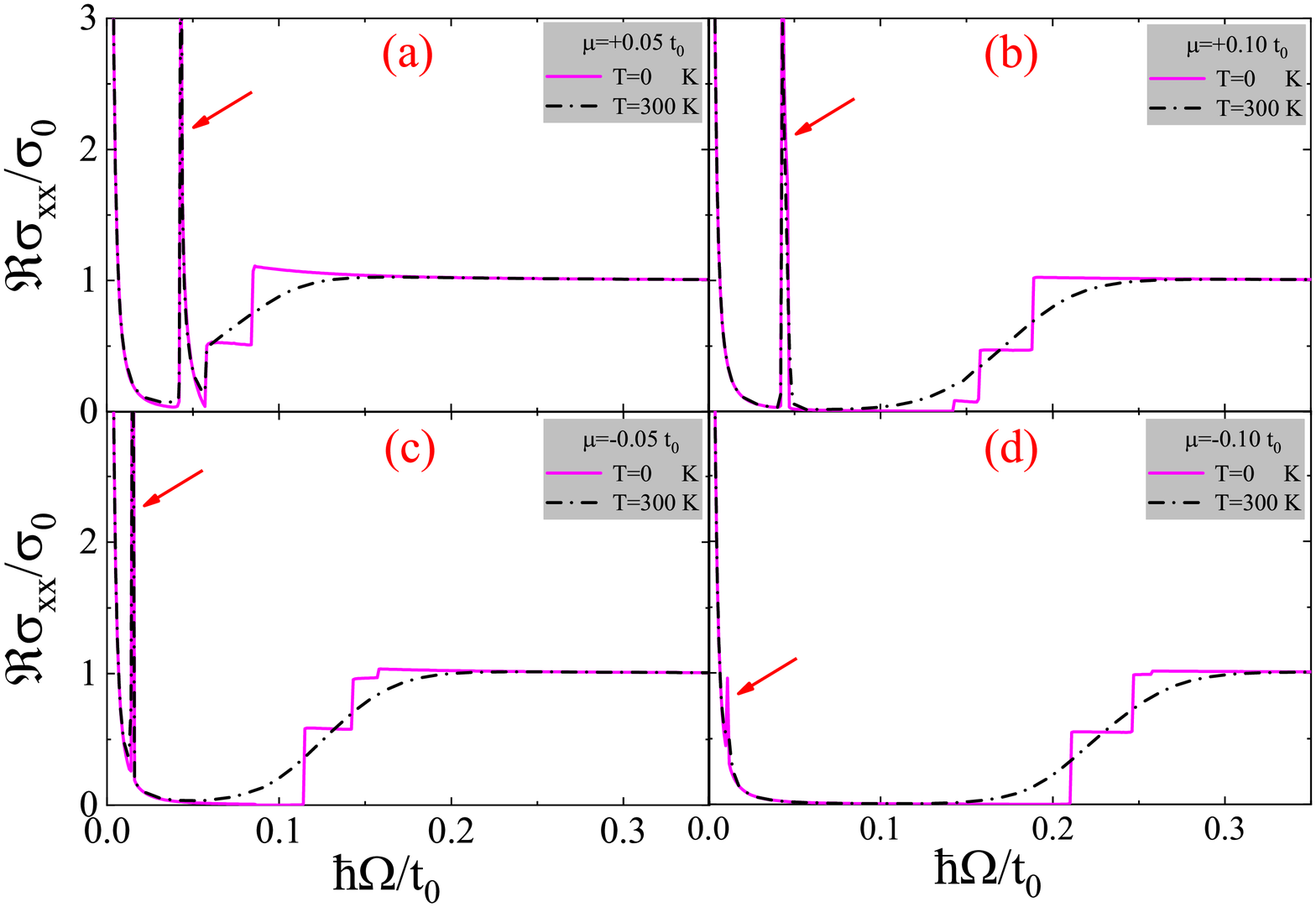}
\caption{The optical conductivity of Kekul\'{e}-patterned graphene for the case of $\delta t=0.1t_{0}$ and $u=0.1t_{0}$ at different temperatures, $T=0~K$ (magenta solid curves) and $T=300~K$ (black dashed-dotted curves), and for different chemical potentials (a) $\mu=+0.05~t_{0}$, (b) $\mu=+0.05~t_{0}$, (c) $\mu=+0.05~t_{0}$ and (d) $\mu=+0.05~t_{0}$. It is evident that the band nesting resonance identicated by red arrow, except when the chemical potential approaches the intersection of $E_{1\mathbf{k}}$ and $E_{3\mathbf{k}}$ energy bands ((d) panel), is robust with respect to increasing temperature.}
\label{Fig07}
\end{center}
\end{figure}

Finally it is desirable to mention to the effect of the next-nearest hopping energy deviations on the optical conductivity of Y-shaped Kekul\'{e}-patterned graphene. In Ref. \cite{Peres1} the effect of next-nearest hopping energy on the optical conductivity of pristine graphene was investigated and it was found that its effect is negligible at low energies. Moreover, recently the effect of the next nearest neighbor hopping energy deviation on the electronic structure of kekule-patterned graphene has been investigated analytically\cite{Andrade3}. Comparing equation 14c of this article with our result for the effective low energy Hamiltonian, Eq. \ref{eq10}, shows that including this effect in our calculations only leads to the normalization of $U$. Therefore, it can be expected that including the next nearest hopping energy deviation does not change our results for the optical conductivity in general.

\section{Summary and conclusions}
\label{sec04}

In summary, we investigated the effects of a uniform Y-shaped Kekul\'{e} distortion of C-C bonds on the electronic band structure and optical conductivity of graphene. First we introduce our tight-binding model in which the effects of the Y-shaped Kekul\'{e} distortion was taken into account by including both on-site and hopping energy deviations in the minimal tight-binding Hamiltonian of graphene. Then by projecting the high energy bands onto the subspace defined by the low energy bands we derived a low-energy effective Hamiltonian which was found to be in excellent agreement with one calculated from a diagonalization of the full tight-binding Hamiltonian. It was shown that in addition the coupling of energy bands in different valleys caused by both on-site and nearest-neighbor hopping energy deviations, the other main effect of the on-site energy deviation on the low-energy band structure is that a set of bands gains an effective mass and a shift in energy, thus lifting the degeneracy of the conduction bands at the Dirac point. Then, using Kubo formula, we obtained an analytical expression for the real part of the optical conductivity. In the next section we presented our results for the optical conductivity of as a function of the photon energy. We found that in the zero limit of the chemical potential and the simultaneous presence of on-site and nearest-neighbor hopping energy deviations, the optical conductivity displays a dip-peak structure located at the photon energy corresponding to 2 times the effective hopping energy whose occurrence was explained by considering the allowed optical transitions. This is in contrast to the case of zero hopping energy deviation which similar to pristine graphene leads to a flat interband response at all frequencies. Furthermore it was shown that at finite chemical potential the interband absorbtion edge, which for pristine graphene is at $2\mu$, is splitted into two edges whose splitting depends on the amounts of the on-site and nearest-neighbor hopping energy deviations and also the chemical potential. As a notable result it was found that the Y-shaped Kekul\'{e}-patterned graphene at finite chemical potential also displays a large optical response called band nesting resonance. We showed that this effect is robust with respect to increasing temperature except when the chemical potential approaches the intersection of $E_{1\mathbf{k}}$ and $E_{3\mathbf{k}}$ energy bands. Therefore the occurrence of band nesting resonance is a significant optical signature to detect the Y-shaped Kekul\'{e} distortion in single layer graphene. We also discussed the effect of the next-nearest hopping energy deviations which is found to only lead to renormalization of the effective on-site energy deviation preserving our general results for the optical conductivity.

\nonumber \section{acknowledgment}

This work has been supported by Farhangian University.

%

%
%
%
%
\end{document}